\documentstyle[aps,prl,epsf,twocolumn]{revtex}
\begin{document}
\draft
\twocolumn[\hsize\textwidth\columnwidth\hsize\csname @twocolumnfalse\endcsname
\title{Wave function mapping conditions in Open Quantum Dots structures}
 
\author{M. Mendoza and P. A. Schulz }
 
\address{Instituto de F\'{\i}sica Gleb Wataghin, UNICAMP, Cx.P. 6165, 13083-970,
 Campinas, SP, Brazil}
 
\maketitle
 
\date{today}
 
\begin{abstract}
We discuss the minimal conditions for wave function spectroscopy, in which 
resonant tunneling is the measurement tool. Two systems are addressed: 
resonant tunneling diodes, as a toy model, and open quantum dots. 
 The toy model is used to analyze the crucial 
tunning between the necessary resolution 
 in current-voltage characteristics and the 
breakdown of the  wave functions probing potentials into a level splitting 
characteristic of double quantum wells. 
The present results establish a parameter region where the wavefunction 
spectroscopy by resonant tunneling could be achieved. In the case of 
open quantum dots, a breakdown of the 
mapping condition is related to a change into a double quantum dot structure 
induced by the local probing potential. The analogy between the toy model 
and open quantum dots show that a precise control over shape and extention of
 the potential probes is irrelevant for wave function mapping. Moreover, 
 the present system is a realization of a tunable Fano system in the 
 wave function mapping regime.
\noindent PACS number(s) 73.40.Gk,73.21.Fg,73.21.La
\end{abstract}

%\newpage
%\twocolumn
\vspace*{0.5cm}
]
\narrowtext

\section{Introduction}

Experimental probing of electronic states in systems showing spatial 
quantization  
is probably the most direct visualization of quantum mechanical effects. Such 
probing in condensed matter has been a challenge over decades until the 
development of artificial model structures, initially semiconductor
quantum wells and more recently quasi one (or zero ) dimensional 
mesoscopic systems. The control over the design and fabrication of these 
structures lead naturally to the introduction of well defined
local probes of the electronic states. A landmark in 
the wavefunction spectroscopy is the optical probing of quantum-well 
eigenstates by Marzin and Gerard more then ten years ago 
\cite{marzin}. 
The basic idea introduced in this work is that a very thin barrier, which can 
therefore be considered as a delta function, is grown within the 
quantum well at a certain position, leading to a potential perturbation of the 
form $V\delta (z-z_0)$. Such perturbation probes the probability density 
at $z_0$ 
by means of the eigenvalues, $E_i$, shifts, which in first-order 
  approximation are simply:
  
 \begin{equation}
 E^{'}_i=E_i + V|\Psi_i(z_0)|^2
 \end{equation}
 
 In the work by Marzin and Gerard, these energy shifts were obtained by 
 photoluminescence measurements performed in a set of nominally identical 
 quantum wells but with the perturbative barrier located at different positions. In 
 other words, such mapping rely on measurements 
 performed on different samples, each one probing the wave function at a designed 
 position. Later on, Salis and coworkers 
 \cite{ensslin} 
 performed a wave function spectroscopy
  on a single 
 parabolic quantum well, where the electron distribution was 
 displaced with respect to a fixed perturbative barrier by applying an electric field.
 The energy shifts were obtained now by magnetotransport measurements. The great 
 advantage of this procedure, namely the spectroscopy on a single sample, is 
 somehow eclipsed by the fact that only a specific system (parabolic quantum wells) 
 is suitable for it. A variation of this spectroscopy is the introduction of 
 monolayers with magnetic ions embeded in different positions of a quantum well, 
 using the Zeeman spliting as a probe for the wave function 
 \cite{yang}. 
 An alternative approach, based on energy shifts measured 
 by means of resonant tunneling, has been proposed also a few years ago 
 \cite{latge}. 
 Now 
 the mapping of the probability density along the quantum well is 
 related to shifts of the resonant tunneling current peaks for 
 an ensemble of double barrier tunneling diodes, where each sample has a
  perturbative potential spike 
 located at a specific position. 
 This tunneling
 wavefunction spectroscopy has not yet been experimentaly verified. 
 Nevertheless, magnetotunneling has been used as a tool for imaging of 
 electron wave functions in self-assembled quantum dots 
 \cite{vdovin}. 
 
Imaging of wave functions, in spite of the efforts mentioned above, has experienced 
a growing interest mainly due to the use of scanning probe microscopes in searching 
local electron distributions in mesoscopic systems. Within an already long list of 
achievements, it is worth mentioning the study of Bloch wave functions in quasi one 
dimensional systems, such as single wall carbon nanotubes 
\cite{lemay} 
and imaging of bound states in quantum corrals 
\cite{corral}. 
In both cases scanning tunneling microscopes 
were used. Closely
related to the approaches using perturbative potential spikes are the use of 
atomic force microscopes with the measurement of shifts in the conductance across 
a mesoscopic system as a function of the position of the potential perturbation 
induced by the tip of the AFM. An interesting application of this method is the 
imaging of coherent electron flow from a Quantum Point Contact 
\cite{topinka}.

In the present work we analyse the suitability of such imaging procedure for 
quasi-bound states in open quantum dot system in the resonant tunneling regime.
 It can be 
considered the two-dimensional counterpart of the probing of quasi
 bound states in 
double-barrier quantum wells, considered as a toy model.  We are here mainly 
interested in the conditions that maximize the energy shift of the 
resonances in the transmission probability, without breaking the perturbative 
regime within the mapping of the wave function can be established. 
 In the present situation we are
dealing with the quasi-bound states of a double point contact in the resonant 
tunneling regime, 
a rather different situation than single quantum 
point contacts 
\cite {topinka}, 
theoretically discussed within a similar 
framework \cite {guang}. Although our main concern is the mapping 
of quantum dot states, related to resonance shifts in energy, the analysis 
could also be extended to the behavour of the transmission probability plateaus 
related to the quantum point contact channels 
\cite {rocha}.

An important point in the present work is that, if a wave function mapping 
could be experimentally achieved, the open quantum dot system coupled to an
AFM tip would be a realization of a tunable Fano system. Fano resonances 
have been recently observed in electronic transport through a 
single-electron transistor 
\cite {goeres}, 
but a tunability of the effect has 
been reached only in the presence of magnetic fields 
\cite {kobayashi}, 
with the quantum dot in an Aharonov-Bohm interferometer. The degree of freedom 
introduced by the movable AFM tip opens a new possibility for such tuning in 
the absence of magnetic field effects.  
Although Fano resonances have been discussed before in the 
context of mesoscopic systems, the present work 
proposes a possible experimental realization of former theoretical predictions
\cite {chang}.

\section{Wave function imaging in a toy model}

 The wavefunction mapping in double-barrier 
resonant tunneling devices is our toy model to discuss how far can 
a resonant transmission probability peak be shifted, within a simple approach 
that contains the essential features related to the problem. 
The coherent transmission probability is calculated 
in the effective-mass approximation for a double-barrier  
structure with an embeded perturbative barrier, as a function of electron 
incident energy 
\cite {price}. 
Having in mind $GaAs/Al_xGa_{1-x}As$ 
structures 
\cite{adachi}, 
the double-barrier potential profile, considering the conduction band $\Gamma$ 
minimum, is illustrated in the inset of Fig.1(a). 
The relevant 
parameters are the ratio between barrier heights, $H/V_b$; and the ratio between 
the characteristic widths, $L/L_W$. 
 Examples of 
transmission probabilities as a function of incident electron energy are 
shown in Fig. 1(a).

The use of potential spikes at controled positions 
as a mapping tool for the probability density inside a 
double-barrier resonant tunneling diode has a severe limitation in the 
resolution of the energy shifts obtained from  rather broad 
current-density voltage characteristics peaks. On the other hand, increasing
 the energy shift 
of a quantum well resonance has an intrinsic upper bound. As an example 
for the lowest state, this upper bound is achieved when the energy shift 
 $\Delta E_1=E^{'}_1-E_1$, as a function of spike position $z_0$, 
 is comparable to the energy difference $\Delta E_{12}$, between the lowest 
 two quasi-bound states of 
 the system.

  The evolution of the energy shift is illustrated
   in Fig. 1(a) for a quantum 
well $L_W = 150 \AA$ wide. The probing potential spike is  
at the center of the structure with $H/V_b=1$ and $L \leq 30 \AA$.
The lowest two resonances of a unperturbed 
double-barrier quantum well is set as a reference (thin continuous line). 
Energy shifts due to spikes
one and three monolayers thick (dashed line, $ L=3 \AA$, and long-dashed line ,
 $L = 10 \AA$, respectively) show the same qualitative features. 
Having in mind eq.(1), we see that the shift of the second resonance should be 
zero. This shift, however, is non zero and negative 
due to the finite 
thickness of the spike and second order effects
\cite {marzin}. 
On the other hand, the upper limit, 
$\Delta E_1 \approx \Delta E_{12}$ is reached for 
$L = 30 \AA$ (approximately 10 monolayers, thick continuous line) or $L/L_W=0.2$.
 Now, the resonances correspond to a double quantum well, 
where each well is $L^{'}_W=60 \AA$ thick. 

An example of wavefunction mapping for the lowest and second quasi-bound state,
given by the energy shift $\Delta E_i=E^{'}_i-E_i$, as a function of the 
probe potential spike position, is shown in Fig.(1b) for the thick
 probe potentials case, $L/L_W=0.2$. Two aspects are relevant:(i) the
  probability density mapping is  possible for thick potential probes, 
as far as the potential heigth is below a critical value; and (ii), above a
critical potential height the energy shifts,
 $\Delta E_i$, as a function of spike position $z_0$ show pronounced 
 singularities, related the fact that $\Delta E_i$ 
 is comparable to $\Delta E_{12}$. Therefore, 
 the mapping of the envelope wave function is restricted to situations 
 where $\Delta E_i < \Delta E_{12}$, i.e., to the left of the crossover shown
 in the example of Fig.2: for 
 $H=V_b$ and $L_W=150 \AA$, this crossover occurs at 
 $L \approx 12 \AA$, indicating an upper limit, $\Delta E_1 \approx 18 meV$ 
 for the energy shifts that still can be associated to a reliable 
 wavefunction mapping.
 
 A diagram indicating a 
  parameter region for such reliable mapping for 
  the lowest state is given in the inset of Fig.2. 
  Here, $\Delta E_i$ ,at the crossover 
  described 
  in Fig. 2, is depicted as a function of a normalized perturbation strength, 
  $HL/L_W$. The 
   appropriate parameter region for a wavefunction mapping
    is the one below the straight line in the figure. This 
  linear behaviour indicates a scalling of the energy shifts with the 
  perturbation strength. Energy shifts up to $\Delta E_i \approx 35 meV$
   can be achieved, which could be resolved in experimental I-V
   characteristics of usual double-barrier diodes.

However, the main point from such a toy model calculation  
is that wide perturbative 
spikes, up to $L/L_W= 0.2$, still lead to reliable  
 wave function imaging, an important generalization of  eq.(1).
  This one dimensional result help to 
understand that extense potential bumps (provided that they are low enough) 
induced by AFM, indeed probe the wave functions in mesoscopic systems. 
In what follows we 
will be able to extend this result in a  simulation of the wave 
function mapping inside an open quantum dot. 
 
\section{Imaging of Wave functions in open quantum dots}

\subsection{Model calculation}

The transmission probabilities through an open quantum dot are calculated 
within a Green's function formalism applied to a lattice model in the 
tight-binding approximation. This method has already been described throughout 
the literature and has been applied in a variety of problems in the context of 
mesoscopic systems 
\cite{ferry,sols,datta}. 
For the sake of clarity this method is briefly sketched below. 
 
 The open quantum dot structure, emulated by a tight-binding lattice model is 
depicted in Fig.3(a). The black circles represent the lattice sites that define 
a square quantum dot conected to two dimensional contacts 
to the left and to the right by point contacts. The size of the 
quantum dot is $S_{QD}=15a \times 15a$, where $a$ is the host lattice parameter. 
The circles inside a square represent 
a potential column simulating the perturbation induced, for instance, by an 
AFM tip located on the sample at that position. In what follows we consider 
perturbations of a single host lattice site, which corresponds to a extension 
relative to the quantum dot are of $S_P \approx 4.5\times10^{-3}S_{QD}$, up to a 
$5 \times 5$ column, corresponding a relative extension of 
$S_P \approx 0.1S_{QD}$.
 
It should be kept in mind that 
lattice models, with nearest neighbor interactions only, are usually thought as 
simple, although useful, approximations for superlattices or arrays of quantum 
dots, where each quantum well or quantum dot is represented by a site of the 
lattice, respectively. Apart from this extreme lattice limit, lattice models 
are also useful in emulating the bottom of semiconductor conduction bands
 that are 
well described 
by the effective mass approximation. 
In the present work, the
tight-binding hopping parameter is chosen in order to emulate the electronic
effective mass for the GaAs bottom of the conduction band, $m^*=0.067m_0$. 
Since,
$V_{x,y}=-\hbar^2/(2m^*a^2)$, $V_{x,y}=0.142$ eV for a lattice parameter of 
$a=20$ \AA. 
Such parametrization represents quantum dots with lateral sizes up to 
$L_D=300 \AA$, Fig.3(a), still
an order of 
magnitude lower than the typical dimensions of actual quantum dots constructed 
by litographic methods. 
However, the present results have 
the intention of illustrating the probing of the local probability density and 
the relevant scale is the ratio between the extension of the perturbative 
spike and the dot dimension, $S_P/S_{QD}$. 

The AFM tip can also be seen as a controllable impurity in a quantum 
dot and therefore a simple tunable experimental realization of a multiply 
connected nanostructure 
\cite {yong}. 
 In the present approach, a continuous system is discretized into a 
tight-binding lattice, considering a 
single $s$-like orbital per site and only nearest-neighbour hopping elements. 
These 
two parameters are the only ones necessary for describing the electronic behaviour 
in lateraly modulated heterostructures near the bottom of the GaAs conduction band.
The device region of a Open Quantum dot system modeled this way, Fig. 3(a), 
is $M=45$ sites long and $N=25$ sites wide. 
The total Hamiltonian, $H_T$, is a sum of four terms:
the 
dot and the two point contacts regions, described by the $H_D$, and the left and rigth 
contact regions, $H_L$ and $H_R$, respectively, and the coupling term between 
the contacts and the dot structure, $V$:

\begin{equation}
H_T=H_D+H_L+H_R+V
\end{equation}

We are interested in the transmission, $t_{\nu,\nu^{'}}$, and reflexion, $r_{\nu,\nu^{'}}$, 
amplitudes, related to the $G^+(\nu',r,\nu,l,E)$ and $G^+(\nu',l,\nu,l,E)$  
Green's functions, respectively. Here,
$l(r)$ stands for a sites column at the left(rigth) of the OQD device, as indicated in Fig.3(a);
while $\nu(\nu')$ are transverse incident(scattered) modes in the contacts at a given energy 
$E$. The first step is calculating the Green's functions of the semi infinite contacts, 
$C_L$ and $C_R$:

\begin{equation}
G^+(E)=\sum_{\nu,\mu}\frac{|\psi^{\nu \mu}><\psi^{\nu \mu}|}{E-E^{\nu \mu}+i\eta},
\end{equation}

where $|\psi^{\nu \mu}>$ and $E^{\nu \mu}$ are the eigenstates and eigenvalues of the contact 
regions, with $\nu (\mu)$ as transverse(longitudinal) quantum numbers. Actually, we need the 
matrix elements of the Green's functions for the $l$ and $r$ sites columns, given by:

\begin{equation}
G_{l(r)}(n,n')=\sum_{\nu=1}^N\chi_n^{\nu}(\chi_{n'}^{\nu})^*\frac{e^{i\theta_{\nu}}}{|V_x|};
\end{equation}

with

\begin{equation}
\theta_{\nu}=cos^{-1}[\frac{(E-\epsilon _{\nu})}{2V_x}+1]
\end{equation}

and

\begin{equation}
\chi _{n}^{\nu}=\sqrt{\frac{2}{N+1}}sin(\frac{\pi \nu n}{N+1})
\end{equation}

The device region can be decoupled in $M$ transverse chains with $N$ sites each. 
The Hamiltonian for one of these chains, $i$, is writen as:

\begin{displaymath}
H_i=\sum_{n=1}^{N}(|i,n>\epsilon_{in}<i,n|
\end{displaymath}
\begin{equation}
+|i,n>V_{n,n+1}<i,n+1|+|i,n>V_{n,n-1}<i,n-1|),
\end{equation}
where the hopping elements at the edges are $V_{N,N+1} = V_{1,0} = 0$. The corresponding 
Green's function is:

\begin{equation}
G_i=[(E+i\eta ){\bf I}-H_i]^{-1}
\end{equation}

The Green's functions $G^+(\nu ',r,\nu ,l,E)$ and $G^+(\nu ',l,\nu ,l,E)$ 
are calculated 
by means of a recursive procedure, coupling the Green's functions of successive transversal 
chains along the device, eq.(8), based on the Dyson equation

\begin{equation}
 G=G_0+G_0VG=G_0+GVG_0
\end{equation}

The starting point of this iterative procedure is the Green's function,
 given by eq.(4), corresponding to a transversal chain 
  at the right, ${\it r}= M+1$ $(G_r)$, of the open quantum dot 
 structure, successively coupled to the device chains, $G_i$, and finally to 
 the left contact, $G_l$.

The transmited and reflected amplitudes are:

\begin{equation}
t_{\nu \nu'}(E)=i2|V_x|\sqrt{sin\theta_{\nu'}sin\theta_{\nu}}e^{i(\theta_{\nu l}-\theta_{\nu ' r})}G^+(\nu ',r,\nu ,l,E)
\end{equation}

and

\begin{displaymath}
r_{\nu \nu'}(E)=i\sqrt{\frac{sin\theta_{\nu'}}{sin\theta_{\nu}}}e^{i(\theta_{\nu}+\theta_{\nu '})l}
\end{displaymath}
\begin{equation}
\times[2|V_x|sin\theta_{\nu}G^+(\nu ',l,\nu ,l,E)+i\delta_{\nu ' \nu}]
\end{equation}

The total transmission probability, the quantity discussed in what follows, is given by the 
Landauer-B\"uttiker formula:

\begin{equation}
T(E)=\sum_{\nu '}^{N}(\sum_{\nu}^{N}|t_{\nu '\nu}(E)|^2)
\end{equation}

\subsection{Numerical Results: energy shifts and imaging}

The main limitations of resonant tunneling mapping of the wave function, 
namely the broadness of measuread I-V characteristics, as well as the 
uncertainties related with a procedure involving a set of different samples, can 
be overcomed in the imaging of quasi-bound states in open quantum dots. The 
embeded potential spikes are substituted by the potential bumps induced by a
AFM tip scanned over a single sample and the resonant tunneling current, a 
rather wide integration of transmission probability resonances, is reduced to 
single and well defined conductance peaks. Although imaging of coherent electron 
flow through a quantum point contact has been reported 
\cite {topinka}, 
where the mapping is achieved by measuring deviations of the quantized conductance plateaus as a 
function of AFM tip position, it remains to be properly discussed the use of 
energy shifts of conductance peaks to image the wave function inside a quantum 
dot.

Typical transmission probabilities as function of 
incident energy are shown in Fig.3(b). Here we clearly see two resonances due 
to quasi-bound states in the quantum dot below the threshold of the first 
quantized conductance plateau due to the quantum point contacts that connect 
the dot to the left and right two-dimensional reservoirs. The thin continuous line 
is for the unperturbed quantum dot. The other curves are for a potential bumps 
at the center of the dot with $H=0.05 eV$, but different sizes. It should be 
noticed that this is 
actually a strong perturbation, since the energy separation between the two 
resonances in the bare dot is $\approx 0.01 eV$. The dashed line is for a 
delta function like bump, with $L=1$. It can be seen that a small shift occurs 
for the lowest resonance, while the second one remains unchanged as expexted. 
The long dashed curve is for a wider bump, $L=3$, with  corresponding larger 
shifts of the resonances. The thick continuous line is for $L=5$ revealing the 
signature of a 
doublet resonance of a symmetrically 
structured dot, instead of slightly single quantum 
dot perturbed levels. A clear analogy to the double barrier 
structure, Fig. 1(a), can be established.

The mapping of the probability density is obtained by scanning the potential 
bump across the quantum dot in both directions. This procedure introduces 
asymmetries in the structure as far as the perturbation is not at center of 
the structure, but the figure of merit is the position in energy of the 
transmission resonances and not the peak heigths. For the strength of the 
perturbation in the results shown in Fig. 3(b), the mentioned analogy with 
the results in Fig. 1(a) should be taken carefully. Indeed, such a high 
perturbation potential, $H = 50 meV$, strongly affects the transmission 
channels when placed near the quantum point contacts. This is illustrated in 
Fig. 4, where the energy shifts of the lowest and second resonances are 
depicted as a function of the position of two different perturbative bumps. 
Fig. 4(a) represents a bona fide mapping of the probability densities for 
a very low, although spatially extended, perturbation: $H = 5 meV$ and $L = 5a$; 
 while Fig. 4(b) shows a inadequate mapping for $H = 50 meV$ and $L = 3a$. 
 The cusps in Fig. 4(b) are artifacts due to mode couplings and show no 
 resemblance with the actual shapes of probability densities maxima, while the 
 behavior of the energy shifts in Fig. 4(a) are qualitatively in agreement 
 with the probability densities for the two lowest states of the unperturbed 
 system.
 
 The differences between a fair and a inadequate mapping situations become 
 clearer by looking at the contour plots of the energy shifts as a function of 
 the probing potential position, Fig. 5, for the same cases 
 shown in Fig. 4. In Fig. 5(a) we see a fair mapping for quasi bound 
 states in an open quantum dot with a high probability 
 density leaking into the quantum point contacts. This is not the case in 
 Fig. 5(b), where the heigh of the potential bump, positioned near the 
 quantum point contacts, strongly suppresses the 
 resonant tunneling channels, turning the open system into a closed one. An 
 appropriate mapping is also obtained for a even wider, $L=7a$,
  low potential bump 
 ($H= 5 meV$) (not shown here). The interesting point here is that the lateral 
 size of the perturbative bump is almost the half of the lateral size of the 
 quantum dot been probed, corresponding to a bump to dot areas ratio of 
 $S_P \approx 0.2S_{QD}$. Therefore, also for a two dimensional probability 
 density  mapping, the 
 upper limit for the spatial extension of the probing potential is not crucial,
 as far as the corresponding heigh of the potential is kept low enough. 
 
\subsection{Tunable Fano resonances}

As pointed out in the introduction, if a wave function mapping 
could be experimentally achieved, the open quantum dot system coupled to an
AFM tip would be a realization of a tunable Fano system. Fano resonances 
have been  observed in electronic transport through a 
quantum dot 
\cite {goeres}, 
but a completely tunable resonance has 
been reached only with 
the quantum dot in an Aharonov-Bohm interferometer 
\cite {kobayashi}. 
The variation of the connecting channels, achievable by changing gate voltages may 
provide a partial tunability 
\cite {goeres}, but an extra degree of freedom, 
introduced by the movable AFM tip, allows such tuning in 
the absence of magnetic field effects.  

Asymmetric Fano line shapes are the result of the interference between a 
resonant and non-resonant scattering paths. For weakly coupled states, the 
line shapes of the associated resonances are Lorentzian like. This is the 
case of the resonances shown in Fig. 3(b). However, when such a resonance 
occurs at energies near the onset of a conductance plateau, the line shape 
of the transmision resonance may change to a Fano like one. 
This changing the resonance line shape is illustrated in Fig.6, for a similar 
open quantum dot as shown previously
(now $S_{QD}=7a \times 7a$). 
 Fig. 6(a) and Fig. 6(b) are for perturbative
bumps at the positions indicated in the insets. It can be seen that the third 
resonance occurs at an energy where the transmission probability through the 
point contacts can not be neglected as in the case for the lower ones. Such 
contribution can be changed with gate voltages that tune the conection between
the dot and the 2D reservoirs 
\cite{goeres,xiong}, 
with a  corresponding modification of the resonance line shape. 

Fig. 6 illustrates how
an extra degree of fredom, introduced by the potential spike, keeping the 
quantum point contacts fixed, leads to the change from a Lorentzian to  a 
Fano-like resonance. We believe that such an extra degree of freedom may permit 
a complete tunability of Fano resonances in open quantum dot sytems in 
absence of magnetic fields.

\section{Final remarks}

The present work addresses the modeling of wave functions imaging by means of 
experimental perturbative approaches. The spectroscopy proposed is based on 
resonant tunneling. First we analyse a one dimensional problem, a resonant 
tunneling diode, closely related to the initial experimental 
proposals 
\cite{marzin}, 
relying on multiple samples experiments. The second 
situation studied here concerns a two dimensional problem, namely an 
open quantum dot in the resonant tunneling regime. 

A remarkable analogy between both situations is established, 
 with an important common result: wavefunction 
mapping is achievable with rather spatially extended perturbative potentials. 
This is in oposition to the initial supositions of delta like perturbative 
spikes 
\cite {marzin,ensslin}, 
but provide a strong support to the imaging using AFM induced perturbations, where the 
exact form and extension of the depletion underneath the tip are not so clearly
controled. We believe that our results open new possibilities to the 
imaging experiments carried out so far on single quantum point 
contacts 
\cite{topinka}. 
An open quantum dot coupled to the tip of an AFM could
also be a new realization of tunable resonance line shapes of 
the conductance through mesoscopic systems. 

\section{acknowledgments}
M. Mendoza would like to acknowledge the Brazilian agency CAPES for financial 
support, while P.A.S. is grateful to the continuous support provided by FAPESP.

\begin{figure}
\caption{(a) Examples of transmission probabilities, as a function of electron 
incident energy, for the structure shown in the inset with $z=L_W/2$, $H=V_b$; 
and $L=0$ (continuos line), $L=3 \AA$ 
(dashed line), $L= 10 \AA$ (long dashed), and $L=30 \AA$ 
(thick continuos line). Inset: double-barrier potential profile with an embeded perturbative 
barrier. (b) Energy shifts for the first (left) and second (rigth)
 resonances of the 
structure in (a) as a function of the position, $z$, of the 
perturbative barrier for $L=30 \AA$, for three 
different perturbative barrier height: $H= 0.043$ eV (lower curves), 
$H=0.172$ eV (intermediate curves) and $H=0.674$ eV (upper curves).}
\label{1}
\end{figure}

\begin{figure}
\caption{Energy shift, $E^{'}_1-E_1$, of the lowest quasi-bound state ; and 
the energy separation, $E_2-E_1$, between the 
lowest and second bound states in a 
double-barrier structure, as a function of the perturbative barrier 
thickness, $L$. The perturbative barrier is located at the center of the well 
with $H=V_b$. The other parameters are the same as in Fig. 1. Inset:
  energy shift $\Delta E = E^{'}_1-E_1$ at crossovers, 
 as a function of the perturbation strength, $HL/L_W$.}
\label{2}
\end{figure}

\begin{figure}
\caption{(a) Schematic illustration of the open quantum dot structure. (b) 
Total transmission probabilities as function of incident energy for the 
structure in (a): bare structure (thin solid line), with a 
potential bump at the center of the structure with $H=0.05$ eV and $L=1a$ 
(dashed line), $L=3a$ (long dashed line) and $L=5a$ (thick solid line).
}
\label{3}
\end{figure}

\begin{figure}
\caption{Energy shifts of the lowest (left) and second (rigth) quasi bound 
states  as a function of the position of the 
potential bump inside the open quantum dot structure. (a) Bona fide probability
density mapping for a wide and low probe potential: $H=5$ meV and $L=5a$. 
(b) Unrealistic mapping for a high probe potential: $H= 50$ meV and $L=3a$.}
\label{4}
\end{figure}

\begin{figure}
\caption{Contour plots of energy shifts, corresponding to the 
situations depicted in Fig.4. The
structure probed is an open quantum dot one and in (a), corresponding to a 
bona fide mapping, the contours indicate finite probability density in the 
contact regions. (b) high probe potentials isolate the quantum dot.}
\label{5}
\end{figure}

\begin{figure}
\caption{Tuning of the resonance at the conductance plateau onset 
with varying tip position: (a) Breit-Wigner 
like resonance, and (b), Fano-like resonance.}
\label{6}
\end{figure}


\begin{references}

\bibitem{marzin}
J-Y. Marzin and J-M. G\'erard, Phys. Rev. Lett. {\bf 62}, 217 (1989).

\bibitem{ensslin}
G. Salis {\it et al.}, Phys. Rev. Lett. {\bf 79}, 5106 (1997).

\bibitem{yang}
G. Yang, J. K. Furdyna, and H. Luo, Phys. Rev. B {\bf 62}, 4226 (2000).

\bibitem{latge}
A. Nogueira and A. Latg\'e, Phys. Rev. B {\bf 57}, 1649 (1998).

\bibitem{vdovin}
E. E. Vdovin {\it et al.}, Science {\bf 290}, 122 (2000).

\bibitem{lemay}
 S. G. Lemay {\it et al.}, NATURE {\bf 412}, 617 (2001).
 
\bibitem{corral}
M. F. Crommie {\it et al.}, Science {\bf 262}, 218 (1993).

\bibitem{topinka}
M. A. Topinka {\it et al.}, Science {\bf 289}, 2323 (2000).

\bibitem{guang}
G-P. He, S-L. Zhu, and Z. D. Wang, Phys. Rev. B {\bf 65}, 205321 (2002).

\bibitem{rocha}
A. R. Rocha and J. A. Brum, Braz. J. Phys. {\bf 32-2A}, 296 (2002).

\bibitem{goeres}
J. G\"ores, D. Goldhaber-Gordon, S. Heemeyer, M. A. Kastner, H. Shtrikman, 
D. Mahalu, and U. Meirav, Phys. Rev. B {\bf 62}, 2188 (2000).

\bibitem{kobayashi}
K. Kobayashi, H. Aikawa, S. Katsumoto, and Y. Iye, Phys. Rev. Lett. {\bf 88}, 
256806 (2002).

\bibitem{chang}
C. S. Kim, A. M. Satanin, Y. S. Joe, and R. M. Cosby, Phys. Rev. B {\bf 60}, 
10962 (1999).

\bibitem{price}
P. J. Price, Superlattices Microstruc. {\bf 2}, 213 (1986).
These results can also be obtained analytically for the 
present situation of zero bias, see H. Yamamoto and X. B. Zhao, Phys. Stat. 
Sol. (b) {\bf 217}, 793 (2000).

\bibitem{adachi}
S. Adachi, J. Appl. Phys. {\bf 58}, R1 (1985).
$\Gamma - X$ coulping has not to be taken 
into account \cite {rossmanith}, since 
the $X$ minimum related quasi-bound state are well above 
the states of interest.
The 
double-barrier potential profile, considering the conduction band $\Gamma$ 
minimum ,is illustrated in the inset of Fig.1(a). The
 effective 
masses are given by $m^{*}=(0.067+0.083x)m_0$ \cite{adachi} 
and conduction-band offsets by $V_b=860x$ meV \cite{rossmanith}. 
\bibitem{rossmanith}
M. Rossmanith {\it et al.}, Phys. Rev. B {\bf 44}, 3168 (1991).

\bibitem{yong}
Y. S. Joe, R. M. Cosby, M. W. C. Dharma-Wardana, and S. E. Ulloa, J. Appl. Phys. {\bf 76}, 
4676 (1994).

\bibitem{ferry}
David K. Ferry and Stephen M. Goodnick, {\it Transport in Nanostructures} 
(Campbridge University Press, 1997), p. 156. 

\bibitem{sols}
F. Sols, M. Macucci, U. Ravaioli, and K. Hess, J. Appl. Phys. {\bf 66}, 3892 (1989).

\bibitem{datta}
S. Datta, Superlattices and Microstructures, {\bf 28}, 253 (2000).

\bibitem{xiong}
S-J Xiong and Y. Yin, Phys. Rev. B {\bf 66}, 153315 (2002).
\end{references}
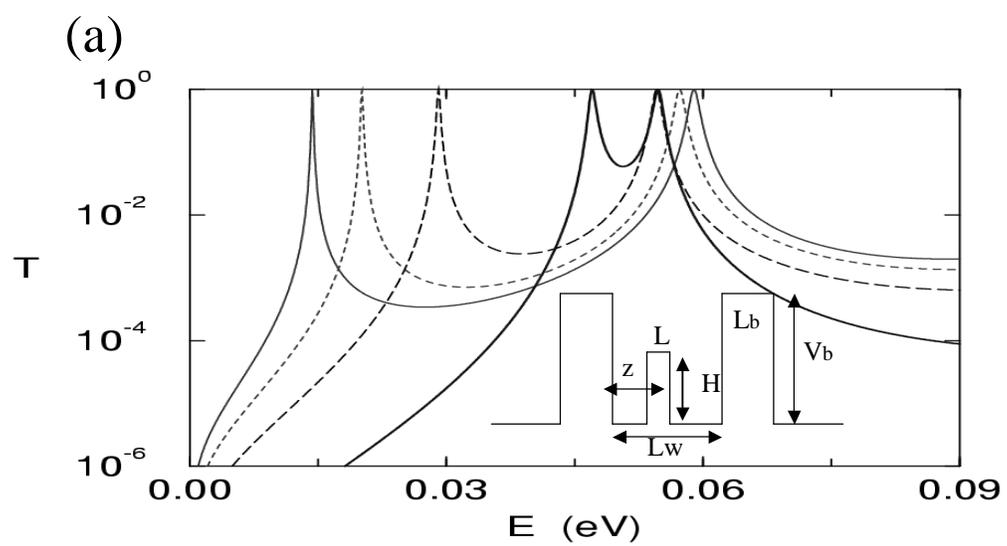
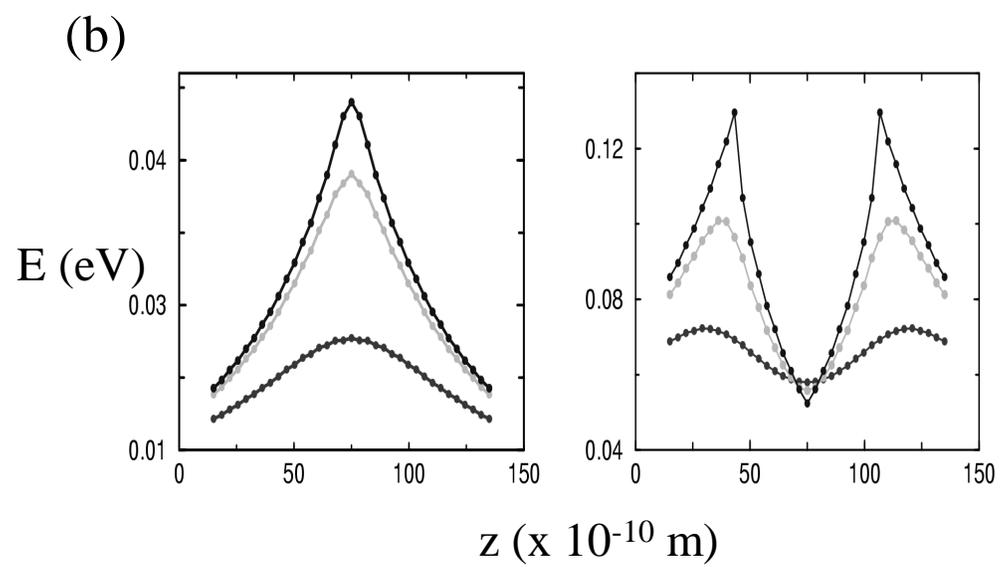
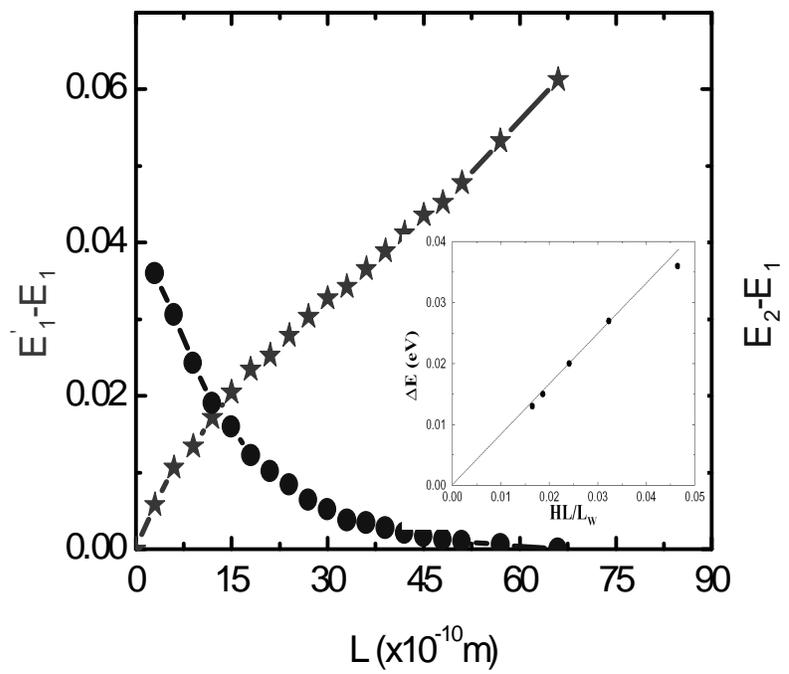
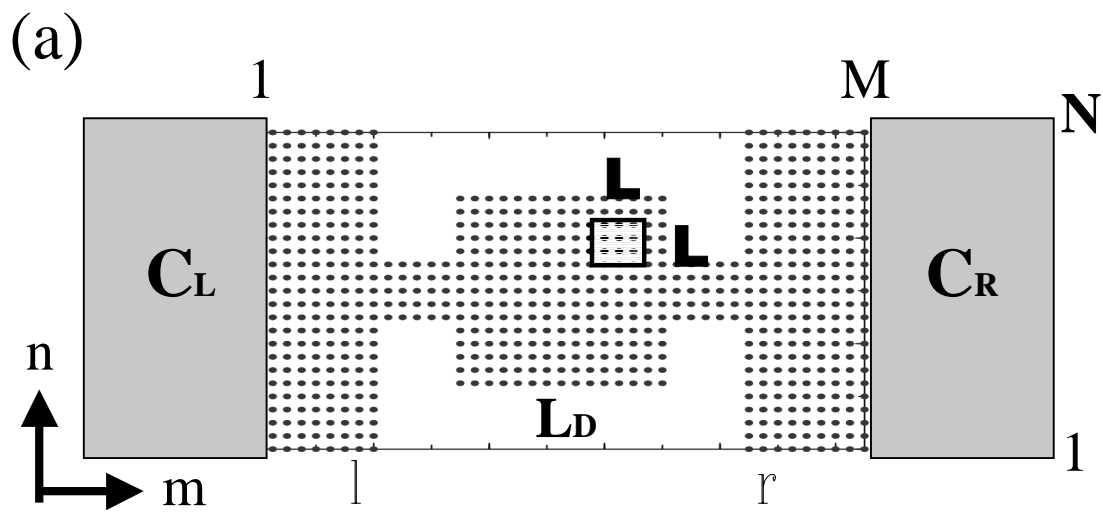
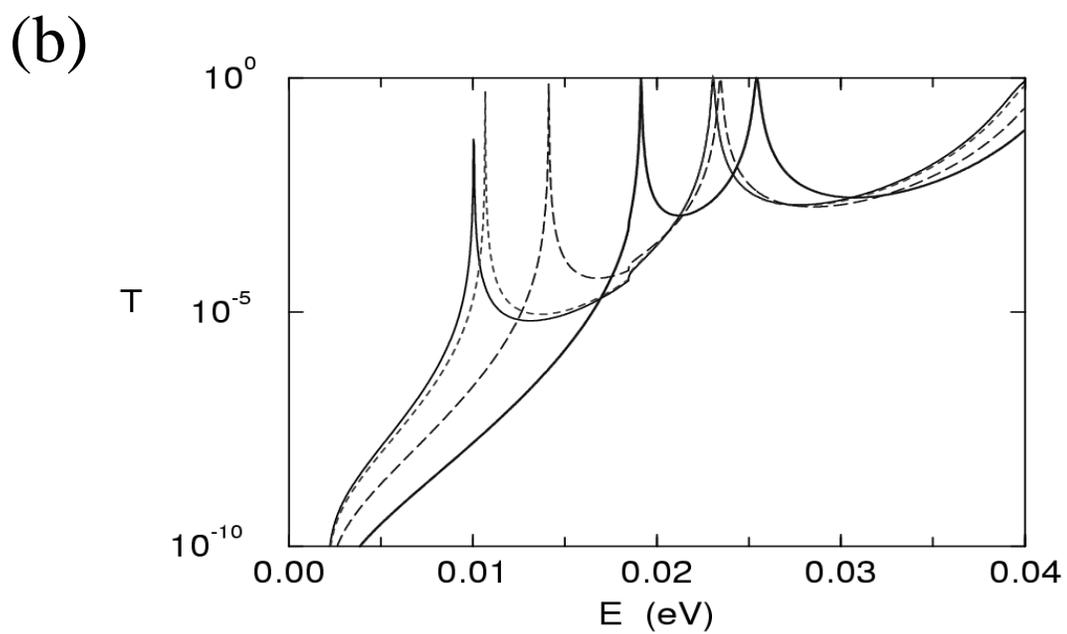
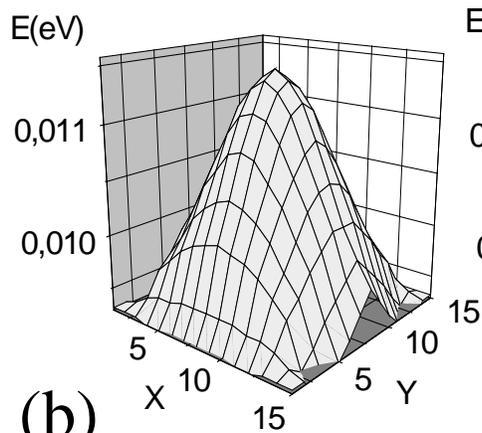
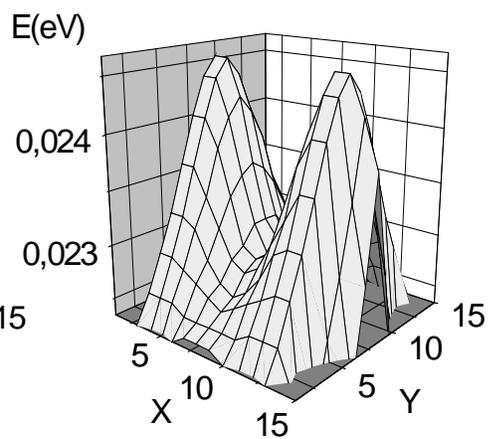
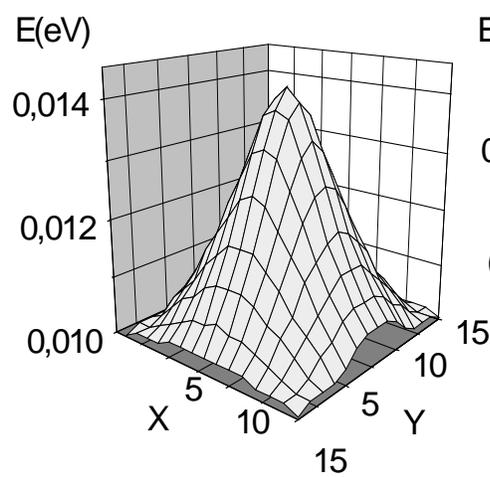
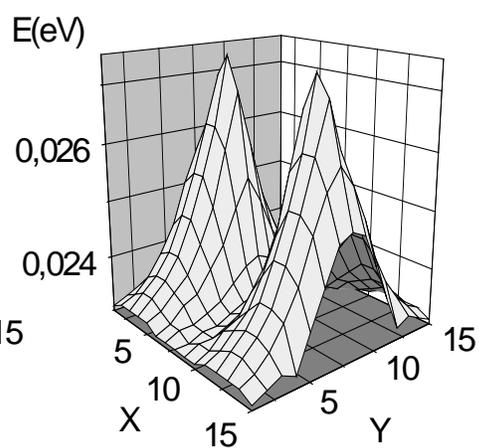
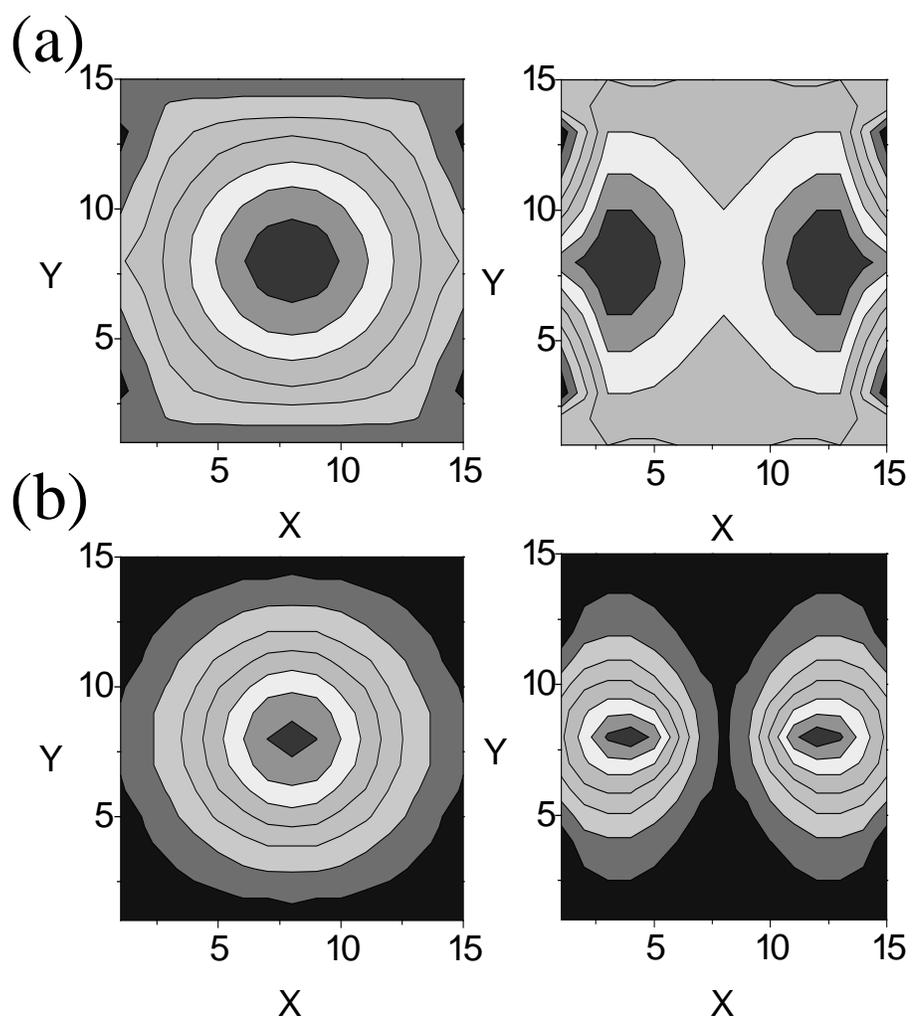
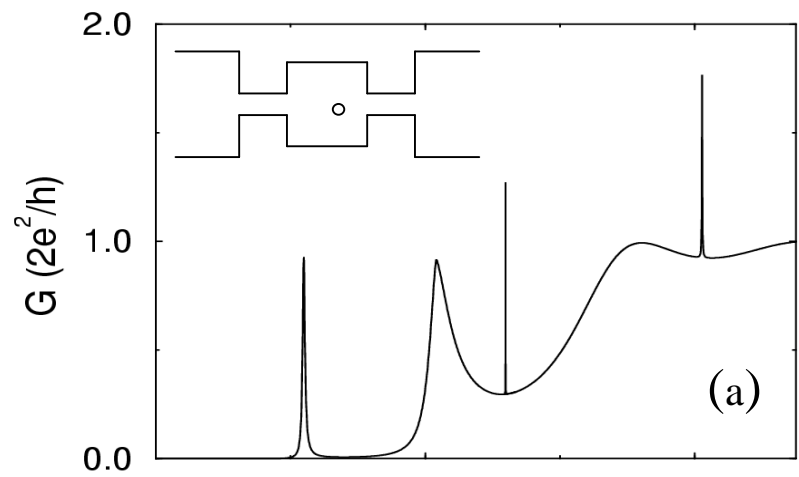
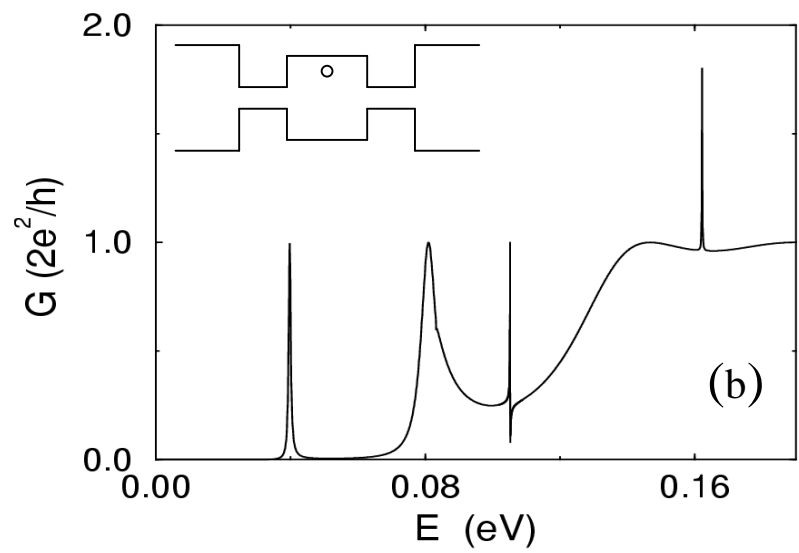
\end{document}